\documentclass[amsmath,amssymb,aps,prd,noshowpacs,nofootinbib]{revtex4}

\usepackage{graphicx}

\usepackage{graphicx}
\usepackage{epsfig}
\usepackage{amsfonts}
\usepackage{amssymb}
\usepackage{amsmath}
\usepackage{bm}
\usepackage{latexsym}

\newcommand{\be}{\begin{equation}}
\newcommand{\ee}{\end{equation}}

\def\fun#1#2{\lower3.6pt\vbox{\baselineskip0pt\lineskip.9pt
\ialign{$\mathsurround=0pt#1\hfil ##\hfil$\crcr#2\crcr\sim\crcr}}}

\newcommand{{\SD}}{\rm SD}

\newcommand{{\Mc}}{\mathcal{M}}

\newcommand{\lan}{\langle}
\newcommand{\ran}{\rangle}

\begin{document}

\title{Magnetic susceptibility  of the quark matter in QCD}
\author{\firstname{Yu.~A.}~\surname{Simonov}}
\email{simonov@itep.ru}
\author{\firstname{V.~D.}~\surname{Orlovsky}}
\email{orlovskii@itep.ru} \affiliation{State Research Center, Institute of
Theoretical and Experimental Physics, Bolshaya Cheremushkinskaya 25, Moscow 117218, Russia}

\date{\today}

\begin{abstract}
\noindent Magnetic susceptibility in the deconfined phase of QCD is calculated
in a closed form using  a recent general expression for the quark gas pressure
in magnetic field. Quark selfenergies are entering the result via Polyakov line
factors and ensure the total paramagnetic effect, increasing with temperature.
A generalized form of magnetic susceptibility in nonzero magnetic field suitable for  experimental and lattice measurements is derived, showing  a
good agreement with available lattice data.
\end{abstract}

\maketitle
\section{Introduction}
A possibility of strong magnetic fields (m.f.) in astrophysics \cite{1,2} as
well as in heavy ion collisions \cite{3,4}, see \cite{5} for a review, poses an
important question: how the QCD matter react to m.f. and, in particular,
whether it is paramagnetic or diamagnetic.

This topic has caused a vivid interest in the physical community recently
\cite{6,7,8,9} and the first numerical results have been obtained for  the
 magnetic susceptibility at zero and finite temperature in \cite{6},
 magnetization in \cite{7}, magnetic susceptibility as a function of
 temperature in \cite{8}, and pressure in m.f. at finite temperature \cite{9}.

 In the analytical approach one should derive these results from the quark
 pressure $\bar P_q(B,T)$ in m.f. $B$ and temperature $T$ in   the deconfined
 phase of QCD for $T>T_c$ and from the corresponding hadron pressure in the
 confining region.

The magnetic contribution to the quark pressure was considered  mostly in the
framework of effective field theories \cite{9*}. We shall follow the  standard
approximation \cite{10*}, generalized with  inclusion of the vacuum QCD
effects.

 Recently the quark pressure in m.f. $\bar P_q (B, T)$ was calculated in a
 simple closed form, in \cite{10}, where the sum over all Landau levels was
 expressed in terms of modified Bessel functions with correct limits for large
 and small m.f. It is important, that in our approach the effect of  the QCD vacuum enters
 in the form of Polyakov lines, which correct the free quark contribution.
  Moreover in \cite{11} a further analysis of quark mass
 dependence of the transition temperature $T_c(B)$ was done, explaining the
 observed \cite{12} decreasing behavior of $T_c(B)$ for small masses $m_q$.

 It is a purpose of the present paper to study magnetic
 susceptibility (m.s.) of the quark matter as a function of temperature using the
 approach of \cite{10,11}, and to compare the resulting curves with the
 numerical data of \cite{7,8,9}.

 The paper is organized as follows. In the next section we define the basic
 quantities and discuss their dependence on $B,T$ and $m_q$  in the case of zero chemical potential $\mu$.
   In section III a
 detailed comparison with lattice data is done and in the final section a
 summary and prospectives are given.

 \section{General formalism} We consider the quark gas in m.f., where each quark undergoes the
 influence of background color fields, which can be expressed in terms of
 field correlators (FC). The  full thermal theory based on FC was suggested in
 \cite{13} and finally formulated in \cite{14, 15}.

 In  the  deconfined phase the quark pressure in m.f. is \cite{10}

$$ \bar P_q (B, T) = \sum_q P_q (B,T), ~~ e_q \equiv |e_q|,$$

 \be P_q (B,T )= \frac{N_c e_qB T}{\pi^2} \sum^\infty_{n=1} \frac{(-)^{n+1}}{n}
 L^n \sum_{n_\bot,\sigma} \varepsilon^\sigma_{n_\bot} K_1
 \left(\frac{n\varepsilon^\sigma_{n_\bot}}{T}\right),\label{1}\ee
 where \be \varepsilon^\sigma_{n_\bot}=\sqrt{m^2_q + e_q B(2n_\bot
 +1-\sigma)}.\label{2}\ee

It is expressed in terms of Polyakov loops $L(T)$, which contain the FC
contribution \cite{14,15}, namely,  when one neglects the bound $q\bar q$
pairs, appearing close to $T_c$, then one can take into account only the large
distance term $V_1(\infty, T)$, calculated via FC in \cite{15, 16},  and the
fundamental Polyakov loop  in this approximation (called in \cite{14} the Single Line Approximation) is
\be
L(T) \equiv L^{(V)}(T) = \exp \left( - \frac{V_1(\infty, T)}{2T} \right),\label{3}
\ee
where $V_1 (\infty, T) $ was found in   \cite{15}, \cite{16}  from the field correlators.
Note, that $V_1 (\infty, T)$ in $L^{(V)}(T)$ entering (\ref{1}), is actually the static $q\bar{q}$ interaction at large distances, which was measured recently on the lattice \cite{add} to be approximately 0.5 GeV for $T>T_c$. We shall use the form (\ref{3}) with this value $V_1(\infty, T) = 0.5$ GeV in what follows, as well as direct lattice calculations from \cite{16*} for the Polyakov loop
\be
L^{(F)}(T)=\exp \left( - \frac{F_1(\infty, T)}{2T} \right), \label{4}
\ee
where $F_1(\infty, T)$ is the free energy, containing all excitations. It was shown in \cite{14}, that $V_1(\infty, T)>F_1(\infty, T)$ and hence $L^{(F)}(T)>L^{(V)}(T)$.

As it is shown in the appendix of \cite{10}, the sum over $n_\bot,
\sigma$ can be done explicitly in (\ref{1}) with the result
$$ P_q (B) = \frac{N_c e_q BT}{\pi^2} \sum^\infty_{n=1} \frac{(-)^{n+1}}{n}
L^n \left\{ m_q K_1 \left( \frac{nm_q}{T}\right)+\right.$$ \be \left.+
\frac{2T}{n} \frac{e_qB+m_q^2}{e_qB} K_2 \left( \frac{n}{T} \sqrt{e_q B +
m^2_q}\right) - \frac{ne_q B}{12T} K_0\left( \frac{n}{T} \sqrt{m^2_q + e_q
B}\right)\right\}.\label{5}
\ee
Eq. (\ref{5}) gives correct limits of $P_q (B)$
for small and large $B$. The quark pressure (\ref{5}) depends on $B,T$ and
$m_q$. We shall be first of all interested in the region of parameters, when
$eB \ll T$ and $m_q\ll T$, corresponding  to the area studied on the lattice.
In this case one can define  the   magnetic susceptibility $\hat \chi_q (B,T)$
\be
P_q (B,T) - P_q (0, T) = \frac{\hat \chi_q}{2} (e_qB)^2+
O((e_qB)^4).\label{7}
\ee

To proceed one can expand the r.h.s. of (\ref{5}) in the Taylor series in
powers of $(e_qB)$. To this  end one can exploit the relation
\be
K_\nu (z) =
\frac12  \left(\frac{z}{2}\right)^{-\nu} \int^\infty_0 dte^{-t-\frac{z^2}{4t}}
t^{\nu-1}\label{8}
\ee
and obtains
\be
P_q(B,T) -P_q(0,T) = \frac{N_c
(e_qB)^2}{2\pi^2} \sum^\infty_{n=1} (-)^{n+1} L^n \sum_{k=0} \left( \frac{ e_q
Bn}{2Tm_q}\right)^k \frac{(-)^k}{k!} K_k \left(\frac{nm_q}{T}\right) \left(
\frac{1}{(k+1) (k+2)}-\frac16\right).\label{9}
\ee

 Note, that the first two terms
$O((e_q B)^0)$ and $O((e_qB)^1)$ in (\ref{9}) identically vanish as well as the
cubic terms, while the quadratic terms can be written as
\be
\frac{\hat\chi_q}{2} = \frac{N_c}{6 \pi^2 }\sum_{n=1,2,...}  (-)^{n+1} L^n
K_0 \left(\frac{nm_q}{T} \right) .\label{10''''}
\ee
As one can see in
(\ref{10''''}) the quark system retains its paramagnetic nature for any $m_q,
T$ provided the Matsubara series over $n$ is convergent. We shall see however
that in (\ref{10''''}) a strong compensation of different terms in the series
occurs.

Indeed, if $  L  \sim O(1)$ and $ \frac{m_q}{T} <1$, one should keep in
 (\ref{10''''}) the sum over Matsubara frequencies, which yields for the total
 m.s.

\be
\hat \chi (T) = \sum_q  \left( \frac{e_q}{e} \right)^2\hat \chi_q (T) = \frac{ N_c}{3\pi^2} \sum _q \left( \frac{e_q}{e} \right)^2
 J_q ,~~  ~~ J_q \equiv \sum_{n=1,2,...} (-)^{n+1}
 L^n K_0 \left( \frac{nm_q}{T}\right).\label{17'}
\ee

 To find $J_q$ we use the integral representation for $K_0$
\be
 K_0 \left( \frac{ m_q n}{T} \right) = \frac12 \int^\infty_0
 \frac{d\omega}{\omega} e^{-n \left(\frac{m^2_q}{2T^2\omega}  +
 \frac{\omega}{2}\right)},\label{17'''}
\ee
and summing over $n$ in (\ref{17'}) one obtains the following form for $J_q$,
\be
J_q=\frac12 \int^\infty_0
 \frac{dx}{x} \frac{y(x)}{1+y(x)}, ~~ y(x) = L \exp \left( -\frac{1}{x} -
 \frac{m^2_qx}{4T^2}\right). \label{17iv}
\ee

\section{Results and discussion}

 Our resulting formula for   $\hat \chi (T)$ is  given in  (\ref{17'}), where
 the integral $J_q$ is defined in (\ref{17iv}). One can see in (\ref{17iv}) that
 the temperature dependence of $\hat \chi (T)$ is defined mostly by the
 Polyakov line  factor $L(T)$, which grows strongly in the considered  region (see e.g. Fig. 3 of the first ref. in \cite{16*}). In addition there is a weak
 logarithmic dependence from the  upper limit   of integration $\sim \ln
 \frac{T}{m_q}$. One can see qualitatively the same type of behavior of $\hat
 \chi (T)$ in the lattice data of \cite{7,8,9}.
 However, to compare (\ref{17'}) with  lattice results one must take into
 account, that m.f. on the  lattice is quantized and has a minimal value,
 dependent on the lattice size, so that one actually refers to the generalized
 m.s. $\hat \chi (B, T)$,
 \be \frac12 \hat \chi_q (B,T) = \frac{P_q (B,T) - P_q (0,T)}{(e_q B)^2} \equiv
 f \left( \frac{\sqrt{e_q B+ m^2_q}}{T}\right), \label{13}\ee
 where $m^2_q$ enters always as $m^2_q + e_q B$, and therefore one can
 introduce  in  (\ref{17iv}) the effective quark mass
 \be m^2_q (eff) = m^2_q + \lan e_q B\ran_{eff},\label{14}\ee
 where $\lan e_q B\ran_{eff}$ depends  in general on the  experimental setup or lattice configuration. One can
 estimate the minimal value of
$\lan eB\ran_{eff}$, on the lattice, $\lan eB\ran_{min} \approx
\frac{6\pi}{(L_sa)^2}$, which gives for the measurements  in \cite{8}, $\lan
eB\ran_{min} \approx 0.023 $ GeV$^2$, and for those in \cite{7}
 $\lan
eB\ran_{min} \approx 0.005 $ GeV$^2$. Therefore we keep in (\ref{17iv}) $m_q
\to  m_q(eff)$ as   in (\ref{14}) with    $\lan eB\ran_{eff}$ as a fitting
parameter  in the interval $0.005 \div 0.04$ GeV$^2$. As a result  we obtain
two sets of curves in Fig.~1 for $\chi = \frac{4\pi}{137}\hat\chi$, which follow closely the data points; one set,
corresponding to $V_1(\infty, T)=0.5$ GeV gives the best fit for $\lan eB\ran_{eff} \approx 0.025 $ GeV$^2$ for the data of \cite{8}
and $\lan eB\ran_{eff} \approx 0.007 $ GeV$^2$ for the data of  \cite{7}. Another set of curves, corresponding to $L^{(F)}(T)$, taken from lattice data \cite{16*}, gives larger values of effective field $\lan eB\ran_{eff}$, 0.08 and 0.2 GeV$^2$ for the data of \cite{8} and \cite{7} correspondingly. One can see a good agreement of our theoretical predictions and lattice results, note also
a close correspondence of   the  $\lan eB \ran_{min}$  with the fitted values
of  $\lan eB \ran_{eff}$ for the curves on the left graph. This fact shows a usefulness
of our definition of the   $m_q(eff)$ and of our approach in general, where the
main features of the  QCD quark matter are incorporated in  Eq. (\ref{1}),
derived from the quark path integrals in the QCD vacuum and  containing quark
selfenergies in the form of the Polyakov lines.
 \begin{figure}[h]
  \centering
  \includegraphics[width=8cm]{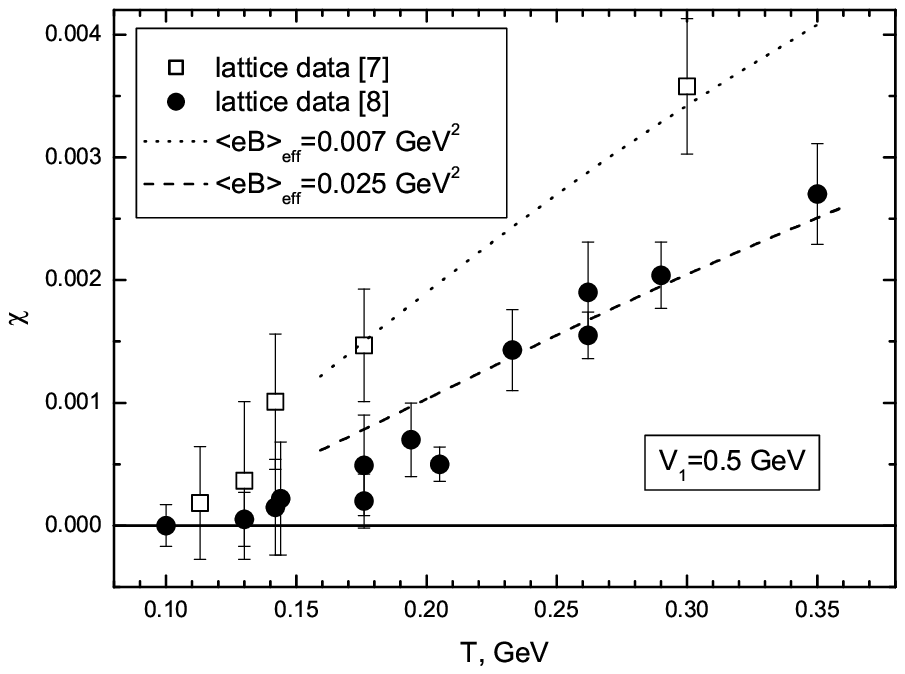}
  \includegraphics[width=8cm]{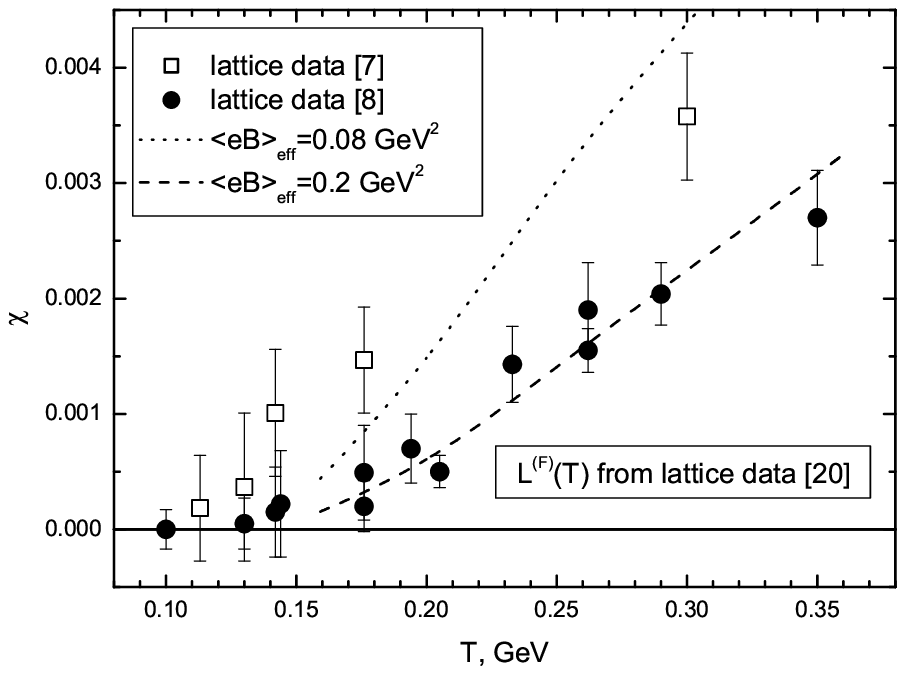}
  \caption{Magnetic susceptibility in SI units ($\chi = \frac{4\pi}{137}\hat\chi$) as a function of temperature for different $L(T)$, a) as obtained from (\ref{3}) with $V_1(\infty, T)=0.5$ GeV (left plot) and b) $L^{(F)}(T)$ from lattice data \cite{16*} (right plot), for different values of $\langle eB \rangle_{eff}$ in comparison with lattice data \cite{7} and \cite{8}.}
\end{figure}

\section{Conclusions}

We have succeeded in obtaining simple formulas for the m.f. dependence of the
pressure and the m.s. of  the quark matter. Our resulting Eq. (\ref{17'})
contains a simple tabulated integral $J_q$. One can see, that $\hat \chi (T)$
strongly depends on $T$ due  mostly to the Polyakov loop and is  almost
insensitive to quark masses when $T\gg m_q$. Both features are supported by the
data of \cite{8}. The resulting magnitude of $\hat \chi(T)$ is strongly reduced by
the  oscillating Matsubara series as compared to the leading $n=1$ term, and is
in a good quantitative agreement with  lattice computations in \cite{7},
\cite{8} and \cite{9}, where the result of \cite{7} is somewhat higher, due to
smaller effective mass values $m_q(eff)$.

A good agreement of our predictions for m.s. with available data is in line
with similar agreement of the transition temperature in m.f. in \cite{10,11},
obtained in the  framework of our theoretical approach \cite{13,14,15,16},
which can be a good starting point for detailed analysis of the quark-hadron
transition.

 The authors are grateful to our colleagues M.~A.~Andreichikov and B.~O.~Kerbikov
 for discussions.

The financial support of the grant RFBR 1402-00395 is gratefully acknowledged.

\end{document}